# The role of the sun in the Pantheon's design and meaning

**Abstract**

Despite being one of the most recognisable buildings from ancient Rome, the Pantheon is poorly understood. While its architecture has been well studied, its function remains uncertain. This paper argues that both the design and the meaning of the Pantheon are in fact dependent upon an understanding of the role of the sun in the building, and of the apotheosised emperor in Roman thought. Supporting evidence is drawn not only from the instruments of time in the form of the roofed spherical sundial, but also from other Imperial monuments, notably Nero's Domus Aurea and Augustus's complex of structures on the Campus Martius – his Ara Pacis, the '*Horologium Augusti*', and his Mausoleum. Hadrian's Mausoleum and potentially part of his Villa at Tivoli are drawn into this argument as correlatives. Ultimately, it is proposed that sun and time were linked architecturally into cosmological signposts for those Romans who could read such things.

## 1. Introduction

The Pantheon is the best preserved architectural monument of the Roman period in Rome (fig. 1). Originally built by Agrippa around 27 BC under Augustus's rule, it was destroyed by fire under Domitian, then rebuilt and finally completed in its present form during Hadrian's reign, in ca. AD 128 (Hetland 2007)[1]. There is a great deal of uncertainty about the original Agrippan design; excavations, however, have suggested that this building was already circular (although probably open to the sky) and orientated in the same direction (Thomas 1997, La Rocca 1999, Wilson Jones 2003: 180–82).

The Pantheon that we can visit today is composed of a rectangular *pronaos* (portico) with three lines of granite columns fronting a circular building designed as a huge hemispherical dome (43.3m in diameter), built over a cylinder of the same diameter and as high as the radius. Therefore, the ideal completion of the upper hemisphere by a hypothetical lower one touches the central point of the floor, directly under the unique source of natural light of the building (fig. 2). This light source is the so-called oculus, a circular opening 8.3m wide on the top of the cupola. It is the only source of direct light since no direct sunlight can enter from the door in the course of the whole year, owing to the northward orientation of the entrance doorway. A characteristic consequence is that the huge mass of the building *usually* (that is, for most of the year) gives to the visitor a strange impression of coldness and dark.

Of the original embellishments the building should have had, the coffered ceiling, part of the marble interiors, the bronze grille over the entrance and the great bronze doors have survived. The interior wall, although circular in plan, is organised into sixteen regularly spaced sectors: the northernmost one contains the entrance door, and then (proceeding in a clockwise direction) pedimented niches and columned recesses alternate with each other. Corresponding to this ground level sector are 14 blind windows in the upper, attic course, just below the offset between the cylinder and the dome (the doorway and large rear niche extend into the attic space so that no blind windows exist there). It is likely that both the niches and the windows were meant for statues, which, however, have not survived (the presence of statues in the Agrippan building is attested by Pliny, *Natural History* 36.38).

As is well known, the Pantheon has exerted a tremendous influence on architecture since the Renaissance (MacDonald 1976). Yet in spite of such a prominent role in history, this building can be connected to other "wonders" of the past, such as the Giza pyramids, in being a monument of a fully literate culture about which practically no written information remains. Indeed, only two Roman sources mention the Pantheon: Pliny, who writes before Hadrian's reconstruction and – as mentioned above - reports only on two statues in it; and the historian Cassius Dio, writing some 70 years after Hadrian, who makes the following cryptic statement: "Perhaps it has this name because, among the statues which embellished it, there were those of many gods, including Mars and Venus; but my own opinion on the origin of the name is that, because of its vaulted roof, it actually resembles the heavens." (Cassius Dio 53.27.2)

As a consequence, we actually do not know why the Pantheon was built and how it was used. The contributions towards solving this problem have analyzed several aspects possibly connected with the symbolism in the project (see, e.g., Loerke 1990, Joost-Gaugier 1998, Wilson Jones 2003, Sperling 2004) and, of course, most authors have more or less emphasised the role of the sun beam (see e.g. Del Monte and Lanciano 1990; Thomas 1997, Rosenbusch in Sperling 1997: 236; Magli 2007: 288-94; Hannah 2009a: 145-56).

It is the aim of the present paper to give an analysis of this issue. Perhaps not surprisingly, as we shall see, this will provide insight also into the broader context of the symbolic and religious world of the Romans in Hadrian's time. In particular, we shall show that the other large-scale project of Hadrian near the Campus Martius area, his Mausoleum across the Tiber, was placed in such a way as to exhibit a symbolic relationship with the Pantheon by means of a hierophany which occurred at the summer solstice.

**2. The sun in the Pantheon**

However the words of Dio may be interpreted, we are "naturally" led to define the Pantheon as a *temple*. Nonetheless, as a temple, the orientation of the Pantheon is unusual. Greek temples were usually orientated – at least in Italy – within the arc of the rising sun, and therefore direct sunlight could enter from the open doors in some periods of the year (Aveni & Romano 1994, 2000); Etruscan temples were orientated "southerly" in the south of east/south of west arc of the compass (Castagnoli 1993) and finally Italic (pre-Roman and early Roman) temples were usually orientated within the arc from the winter solstice sunrise to due south ; therefore direct sunlight could enter every day of the year in all such temples. Furthermore, although a complete analysis of the orientation of the Roman temples in the Imperial period is still awaited, certainly no pattern of orientation to the northern sector of the sky is present.[2] So *why* was the Pantheon oriented to the northern sky? (The monument is actually skewed 5.5 degrees with respect to true north,[3] but this is almost certainly due to a topographical reason as will be discussed in section 4.)

One explanation of the northern orientation is that the project of the building was to some extent inspired by a particular type of sundial, which captured the sunlight within a shadowy interior (Hannah 2009a: 145–54; Hannah 2009b). The type, known from both literature (Vitruvius and Faventinus) and from material remains, was called the *hemicyclium* (Gibbs 1976: 23–27, 194–218; Pattenden 1979). It consisted of a stone block carved out into a hollow hemisphere, with a hole let into its upper surface, through which the sunlight filtered on to the graduated surface inside (fig. 3). Typically these ordinary "roofed spherical" sundials had the sunlight fall on to a series of parabolic lines incised on the dark, interior hemisphere. These lines represent the diurnal passage of the sun, usually on four particular occasions but utilizing just three lines: the summer and winter solstices on separate lines at the extremes of the dial, and (using the same line in between) the two equinoxes. Of course, for this type of sundial to work correctly, the stone face had to be oriented to due south, while the hollow hemisphere was facing north.

That Hadrian's Pantheon can reasonably be compared to such a sundial will become apparent from the technical discussion below, but it actually suffices to mention that after a few visits to the monument on sunny days in different periods of the year, people acquire a (rough) ability to know the day of the year and the hour of the day by looking at the position of the sunbeam. Of course, in being a sort of giant *symbolic* replica of the device, the building must not be confused with a precision instrument. In other words it was not originally conceived and used as an astronomical observatory, although it may have assumed a similar function later on (Sperling 2004). Further, there is nothing beyond the interior moldings of the Pantheon's dome and the marble decoration of its floor which could be recognised as a device used to measure the time with precision, such as markings for the limbs of the sunbeam on particular days.[4] Thus, we stress that all the astronomical analysis which follows does *not* aim to show that the Pantheon was designed to make precise measurements of the sun's cycle, but rather to substantiate the *symbolic* connection of the building with the path of the sun in the course of the year.

To analyze the motion of the spot of sunlight inside the Pantheon we start with a simplified, one dimensional model, fixing the time at (local) noon in the course of the whole year and studying the position of the sunbeam day after day at that specific time. Since the door opens towards the north, the sunlight beam at noon is always located on a "meridian" line which starts from the center of the roof, passing over the entrance and on the wall above it or the floor in front of it (the "meridian" is not precise due to the slight skew of the monument, but the consequences for our calculations are negligible).

For ease of explanation, we start from the autumn equinox. At the autumn equinox, the spot of sunlight touches the interior springing of the upper hemisphere (figs. 4, 5). This is because the sun's altitude coincides with the angle formed by a line connecting the springing with the rim of the oculus. Then at winter solstice – when culmination of the sun reaches its minimum – the spot of sunlight moves up to a maximum height in the roof over the entrance (fig. 6). Thereafter, it moves down, touching again the base of the dome at the spring equinox. In the subsequent days, the beam moves down, illuminating – of course *from inside* - the entrance, which is fully "crossed" around 21 April (figs. 7, 8). After that, the beam starts moving on the floor towards the centre of the building (which of course is *never* reached since the sun does not cross the zenith at the latitude of Rome). From the summer solstice the beam "turns back", re-crossing the entrance between the end of August and the autumn equinox (fig. 9).

From the above discussion, we immediately conclude that the dimension of the oculus was not fixed randomly. Indeed, independently from the dimension of the hemisphere, the angle between the springing and the external rim of the oculus coincides with the altitude of the sun in the days in which the springing will be illuminated. Since, of course, following the sunbeam at noon on different days we are also following the lower boundary of the spot of light on the same days, it could be said that the dimensions of the oculus were chosen in such a way that the sun "spends" autumn and winter in the upper hemisphere of the building.

Immediately after the spring equinox, the beam at noon starts to be visible *from outside* looking through the grille which is mounted over the doors (both doors and grille are with all probability original; certainly they were *in situ* when the monument started to be depicted in the early 16th century drawings). The beam then moves towards the base of the entrance. The entrance is fully illuminated, as mentioned earlier, around 21 April (Gregorian, but in Hadrian's times the delay of the Julian calendar was still minimal, of the order of one day). After this date, the midday spot "enters" the floor.[5]

The above "one-dimensional" approach can now be extended to the complete motion of the sunbeam inside the monument. It is indeed easy to see that every day the sunbeam depicts an arc from west to east, which:
- remains on the upper hemisphere during autumn and winter;
- touches the base of the entrance around 21 April;
- reaches the floor and wanders across it in the central hours of the day from the end of April to the end of August.

It is thus seen that in spring and summer the sun can illuminate the niches and recesses at the base and the blind windows under the springing of the dome. The required altitude of the sun to illuminate the niches/windows is around 56°/60° respectively, and the required azimuth is easily calculated taking into account that each element corresponds to a 22.5° sector of the circular base (360°/16), so that the first niche has an azimuth of 22.5°-5.5° (allowing for the skew from the meridian) = 17°, the second 17°+22.5°=39.5°, the third 39.5°+22.5°=62°, and so on. Using standard astronomical simulation software, which takes into account the variation of the obliquity of the ecliptic (in our case Starry Night Pro 6.0 and Voyager 4.5.4), it can be verified if and when the niches were illuminated (and actually still are, since the variation between then and now is small).

|  | **Altitude of sun 56°** | **Altitude of sun 60°** |
|---|---|---|
| azimuth of niche/window 17° (azimuth of sun 197°) | around 20 April | end of April |
| azimuth of niche/window 39.5° (azimuth of sun 219.5°) | end of April | around 10 May |
| azimuth of niche/window 62° (azimuth of sun 242°) | end of May | around summer solstice |

TABLE 1

As demonstrated in Table 1, only 3 base-niches and only 3 windows (plus the symmetric ones, of course) are ever illuminated inside the Pantheon; all others remain dark during the whole year because the sun fails to be at a sufficient altitude to illuminate them.

All in all, we can say that the architecture of the Pantheon clearly calls attention to some "moments" (actually, short intervals of days): the equinoxes and 21 April. To take the case of the equinoxes: the dimensions of the oculus were fixed; once this was achieved, in order to capture the sunlight as it starts to be visible from outside and finally fully illuminates the entrance, the dimensions of the grille and the very large bronze doors below were fixed accordingly. The disposition (and consequent illumination) of the internal niches is in accordance with this general scheme and, in addition, marks the extreme of the sun's path by signaling the summer solstice.

## 3. The Pantheon and the apotheosised emperor

Roman religion underwent a reassessment aimed to accommodate the divine nature of the emperor exactly in the years of the first building of the Pantheon under Augustus. Thus, since we know that the building was *not* dedicated to a single god, understanding the Pantheon "as a temple" also means trying to understand the way in which the divinization of the emperor was perceived and actualised in Roman religion. It is worth recalling that, according to Dio (53.27.3), Agrippa had originally intended that the building be named after Augustus – presumably as an "Augusteum" rather than a more ambiguous but safer "Caesareum" – as well as containing a statue of the emperor, moves rejected by Augustus in keeping with sensitivities regarding lifetime cult of the emperor within Rome at the time (Wilson Jones 2003: 180; Gradel 2002). The divinization of the deceased ruler was instead, as is well known, established by Augustus at Caesar's death with the help of the appearance of a comet. Caesar's catasterism becomes explicit in Ovid's Metamorphoses, written ca. AD 8, where Caesar's soul "shines as a star" (Ovid, *Metamorphoses* 15.850). Other passages suggest that a particular point of cosmic balance was understood to be assigned to the emperor in the heavens. Vergil (*Georgics* 1. 24-35), somewhat earlier ca. 30 BC, assigns a Caesar to a place in the heavens between Virgo and Scorpius: this is perhaps Julius Caesar also, not Octavian/Augustus, and the location in the sky is where Libra will be newly situated. Manilius (*Astronomica* 4. 546-551, 773-777), writing early in Tiberius's reign, places a Caesar – this time probably Tiberius himself – in Libra (Lewis 2008). Libra, the Balance, at that time at the autumn equinoctial point, is a point of balance between the ecliptic and the equator. Lucan (1.45-59), writing in Nero's time ca. AD 60, has the apotheosised emperor joining the heavens and finding his proper seat on the celestial equator, where he will ensure balance and stability.

Similar concepts are visible "in action" in Nero's Domus Aurea (Oudet 1992, Voisin 1987). The Domus Aurea was a huge palace and effectively the emperor's villa constructed on the Esquiline hill in Rome. Nero's association with the sun is well-attested: on the front of the Domus Aurea complex, facing the end of the Via Appia, there stood a bronze colossus of the emperor in the form of the sun god; and in Nero's later portraits in coin and sculpture he wears the radiate crown usually associated with the Sun god (L'Orange 1947, Hiesinger 1975). Even the strict east-west orientation of the palace has been interpreted as a symbolic representation of the course of the sun across the world, a notion thought to derive from Alexandrian Egypt (Voisin 1987). Most parts of the villa are today lost or buried under successive buildings, but one of the most spectacular rooms still remains: it is an octagonal hall, surmounted by a dome with a central oculus. The room is oriented very precisely in such a way that four sides of the octagon face the four points of the compass. Thus, as in the Pantheon and actually in a more accurate way, the sun at noon points directly to the northern access to the room. Looking at the internal dimensions and the opening of the oculus, it can be seen that:

(a) The position of the north celestial pole defines the perimeter of the oculus. The altitude of the pole always equals the observer's latitude, which is 42°N in Rome. From within this room, it turns out that a person standing at the threshold of any of the doorways of the Octagonal Room can see above the rim of the *oculus* in the dome only that part of the sky with an altitude of about 42° and above. Since the latitude of Rome is 41° 54′ this means that from the vantage point of the southern doorway the observer could just see the area of the sky occupied by the north celestial pole. In effect, therefore, the position of the celestial pole defines the perimeter of the *oculus*. In addition, and perhaps just as significantly, the sun is at about the same altitude (41°) above the horizon at noon on 13 October. This was the anniversary of Nero's accession as emperor in AD 54. At noon on that day the sun was just still visible to an observer standing on the threshold of the northern doorway of the room for the last time until it returned to sight in the first week of March (fig. 10).[6] In between times, the sunlight would fall above eye-level. So from this date of Nero's accession the sun would ascend up the ceiling and then return down again by early March. Thus, Nero's accession combines with the sun and the north celestial pole to govern the dimensions of the oculus.

(b) The summer solstice's midday sun falls completely on the ground within the room, its upper rim practically coinciding with the juncture of the floor and the northern doorway's threshold (fig. 11). So the perimeter of the Room's floor is defined by the sun too.

(c) The lower rim of the equinoctial midday sun strikes the juncture of the floor and the northern doorway's threshold (fig. 12). So the sun measures out the dimensions of this opening, and by extension the other openings, in the walls of the Room. The northern doorway leads to a nymphaeum, so that earth, water and sky are bound together through this doorway (Voisin 1987).

The octagonal room is certainly the key architectural element in the plan of Nero's palace. Its precise function is unknown; perhaps, however, the true meaning of an obscure passage of Suetonius (*Nero* 31), who mentions the existence in the palace of a dinner room in the form of a *rotunda* "rotating day and night"[7] in fact refers to the connection of the room with the celestial cycles.[8]

Nero's Octagonal Room is therefore a convincing case in which architectural elements in a Roman building were symbolically tied to the sun's passages at the equinoxes (and solstices). It is extremely likely that the same occurred in the Pantheon: the springing of the hemispherical dome was thus conceived as an image of the celestial equator. But what about the hierophany that occurred on 21 April (fig. 7)? The days on which the entrance (assumed to be open) is illuminated from the inside are the only days on which the Pantheon does not appear "cold" to the incomers; on the contrary, the huge monument seems to invite the visitors to enter, first with the light filtering through the grille, and then with the full sunlight hitting the entrance area. The month of April was traditionally devoted to Venus, the Goddess from whom the Gens Julia claimed a direct lineage. 21 April is, of course, the traditional date of the foundation of Rome (see, e.g., Ovid, *Fasti* 4: 721-862). Therefore, the symbolic action of the sun on this day is "to put Rome among the Gods". If we suppose, as seems likely, that the emperor was celebrating this precise day there, then his entrance "together with the sun" would have been a symbolic link between the people and the Gods. On the other hand, perhaps connected with the traditions of Italic religion is the relationship between the interior arrangement of the monument, its non-illuminated part and its orientation. As is well known, the city's foundation ritual is described by Roman and Greek historians (such as Varro, Pliny the Elder and Plutarch) as a rule directly inherited from the Etruscans' sacred books (Briquel 2008). A fundamental part of the *libri haruspicini* was connected with the cosmic order, and consisted of the individuation of a terrestrial image of the heavens (*templum*), in which the gods were "ordered" and "oriented". The structure of such "cosmic order" is a complicated and much debated problem, which involves the interpretation of later texts (such as Marziano Cappella) and archaeological remains such as the so-called Piacenza Liver, a 1st century BC bronze model of the liver of a sheep, whose external perimeter is divided into 16 sectors, each devoted to a different god (Aveni and Romano 1994, Rasmussen 2003: 126–48). A "simplified" (8-sector) Roman version is anyway documented archaeologically. It is the augural temple in Bantia, southern Italy, dated to the beginning of the 1st century BC (Torelli 1969; 1995: 97–130). This structure was located on the Bantia acropolis and comprised nine *cippi* (stone cylinders) arranged in a square 3 x 3 array aligned north–south / east–west. In this way the eight directions from the center (which was in itself devoted to the sun) identified the eight main divisions of the cosmos. Each cylinder bears an inscription on the top: the central one is devoted to Sol [SOLEI], the east to Iuppiter [IOVI] and the west to Flusa [FLUS], a local chthonic deity; the others carry information about the quality (bad or good) of the omens brought by birds coming from the respective direction. As was already noticed by Nissen in the 19th century, the internal arrangement of the Pantheon is clearly analogous to the 16-part structure of the *templum* (Nissen 1873, Wilson Jones 2003: 183). The most favorable gods were those of the northern part of the sky, and the northern part of the interior of the Pantheon is the only area which can be illuminated by the sun. In particular, Rome by itself – and, by transposition, the emperor – are "added" to the gods by means of the hierophany occurring at the date of foundation of the town. The "southern" part of the *templum* was devoted to "chthonic" gods, and the southern part of the Pantheon actually remains in the dark during the whole year.

All in all, if the Pantheon has to be considered as Hadrian's *summa* of the relationship of Roman religion with power, it may be suspected that the aforementioned catasterism of Julius Caesar via the comet had a connection with the building as well. Indeed, at least formally, it was the comet which appeared at the time of his funeral which allowed the process of divinisation of the emperor to start. This comet, Pliny informs us (*Natural History* 2.93-94), was taken by the ordinary people to be a sign that the soul of Caesar had been received among the immortal gods. Indeed, Nissen (1873) regarded the Pantheon (of Agrippa) as *the* temple in Rome referred to by Pliny as the sole site in the world of the cult of a comet, and specifically that comet which appeared after Caesar's death during Games that Octavian was celebrating in honour of Venus Genetrix in 44 BC. Nissen (1873: 549-51) also proposed that the unusual orientation of the Pantheon towards the north was a result of a desire to commemorate this comet's appearance over the northern horizon. He was writing twenty years before the excavations in the 1890s, the interpretation of which led to the idea that Agrippa's Pantheon was rectangular and south-facing (Thomas 1997). Instead, Nissen – and now a number of modern archaeologists – took Agrippa's structure to be circular and north-facing. Pliny quotes Augustus himself as noting that the comet was visible for seven days in that region of the sky which is visible *sub septentrionibus*. This may mean literally "under the *Septentriones"*, the name given by the Romans to stars in the constellations of both the Great Bear and, on occasion, the Little Bear, which stand on either side of the invisible north celestial pole (in Roman times, owing to precession, the north celestial pole was not marked by a bright star). Ramsey and Licht (1997: 86-90) argue against such a literal interpretation, at least with regard to the Great Bear, instead preferring a more generalised translation of "in the north", because their calculations of the position of the comet at the time of its appearance make a placement under the Great Bear, i.e. to the north-west, problematic. They do, however, allow the phrase *sub septentrionibus* to be a possible reference to the *Little* Bear, as this would lie between north-north-east and due north at the time of the comet's appearance, the comet itself lying between the constellations of Cassiopeia, Perseus and Andromeda, outside the circumpolar region. It remains impossible, however, to be more precise about the actual alignment of the comet, or to know how the Romans would have noted it for future reference if they wanted to align a structure like the Pantheon towards its rising point.

Be that as it may, it is worth asking what could be observed *from* the Pantheon through the oculus. The oculus leaves only a very small area of sky visible directly above the observer. If we calculate out that circle of sky so as to give a virtual horizon from within the building, and run it astronomically against various dates (equinoxes, solstices, Christian feast days for the building as a church), nothing very remarkable is produced. But if we take observations through the oculus on the supposed dates of the sighting of Caesar's comet in 44 BC – within the period 20-30 July, when Octavian put on the Games for Caesar's triumphs and Venus Genetrix – then we find that the last part of the sky to be visible through the oculus just before sunrise is that band of the sky in which the comet may have been placed.[9] Whether this applies to Agrippa's (circular) Pantheon too, remains, of course, speculative.

## 4. The role of the Pantheon and of the Hadrian's Mausoleum in the "sacred space" of Campus Martius

The area where the Pantheon was first built, the so-called Campus Martius, was originally conceived by Augustus and his architects as a sacred space, where the divine rights of the ruler and his achievements were actualised as tangible monuments on the earth. The area consisted of the following main elements, from south to north (see e.g. Rehak and Younger 2006):

- Agrippa's Pantheon

- a Egyptian obelisk from the Sun temple of Heliopolis (modern north-west Cairo) used as a 30-meter high gnomon. The base of the obelisk (today re-erected in the nearby Piazza di Montecitorio) carries an inscription commemorating the settlement of Egypt by Augustus and dedicating the monument to the sun

- a meridian with the obelisk as its gnomon

- the Ara Pacis, a marble altar celebrating the ruler and his family

- the mausoleum of Augustus, a huge circular building with an octagonal interior plan surmounted by trees.

Also the *ustrinum*, i.e. the place where Augustus's body had to be burned, was certainly located in the area, but its precise location is uncertain.

The bronze meridian line of the "*Horologium Augusti*" has been partially excavated ( Buchner 1982, 1993-94). Buchner attempted to reconstruct the whole project so as to include within its gnomon's real and virtual ambits not only the Ara Pacis to the north-east, but also the Mausoleum of Augustus and the Ustrinum further north. Originally, he proposed that the whole "sundial" consisted of a full "double-axe" network of shadow lines (Buchner 1982), but later he reduced his reconstruction to just a circular "wind-rose" section of this, on the basis of Pliny's testimony (Buchner 1993-94, Pliny, *Natural History* 36.72[10]). A simple meridian line, however, currently seems most likely (Heslin 2007).

Nevertheless, even if the obelisk's shadow was "read" primarily along only a simple meridian line at noon throughout the year, that shadow was still cast elsewhere at other times of the day through the seasons. In Buchner's reconstruction the equinoctial shadow reached within the Ara Pacis, and therefore in his mind the two monuments were intentionally placed so as to obtain a spectacular hierophany. This idea has been criticised as impossible (Schütz 1990), but this criticism seems unwarranted. The obelisk, at almost 30m in height, was able to cast a shadow about 66m long at noon on the winter solstice, to judge from Pliny's assertion that "a stone pavement was laid out in accordance with the height of the obelisk, equal to which was the shadow at the sixth hour on the day of the full winter solstice". At the equinox, the shadow of Augustus' obelisk would theoretically stretch about 89.5m in the direction of the Ara Pacis, which was 83m away. It remains still to be demonstrated that the obelisk's shadow could not reach the Ara Pacis at that distance – the present setting of the obelisk precludes such testing, and Schütz's criticism of Buchner is focussed only on the effective disappearance of the shadow of the *globe* on top of the obelisk, not on the obelisk itself (Schütz 1990). Tests with monuments of a comparable scale suggest that in fact the obelisk's shadow could have reached the Ara Pacis.

The Altar was voted in 13 BC by the Senate, to commemorate Augustus's return from the western provinces of Gaul and Spain, and stood beside the via Flaminia, the road along which Augustus will have traveled on his way back into the city. It was completed in 9 BC, the year the obelisk was also erected. Whether or not the Altar stood at the end of a physical equinoctial line on an enormous net-like grid (a feature cast into some doubt), the general effect at the equinoxes would still hold, as the shadow pointed in the direction of the Ara Pacis, and may indeed have reached it. That effect gains in political and ideological import when we take into account the fact that Augustus's birthday fell within a day or so of the autumn equinox. On his birthday observers could be reminded of Augustus's prime role in bringing peace back to the Roman world after a century of violence, through his settlement of Egypt (the source of the obelisk), and of the western provinces (symbolised by the Altar). At the same time, they would be made aware of his cosmic status, as the sun itself drew his monuments together on his birthday. This cosmological symbolism applies to his conception day too: the northern extremity of the meridian line, which undoubtedly exists, marks the turning point of the noonday sun's shadow at the winter solstice and when the sun entered Capricorn, nine months before Augustus's birthday. Finally, the symbolism was probably incorporated also into the orientation of the Ara Pacis. Indeed, at least judging from existing maps (the monument is not in its original position today) the entrance of the altar was oriented about 18° south of west (Toynbee 1954), so that the opposite side (with a relief probably depicting Venus) was oriented towards the rising sun at the end of April, a month in which Venus was particularly honoured.

Sun and time, then, were linked architecturally into cosmological signposts for those Romans who could read such things. We have long been familiar with this tendency in Augustus's principate from such smaller objects as his coins, minted with the image of Capricorn to denote the astrological sign under which he was born (Lewis 2008). Now we can see it established on a much larger scale. And what applied to Augustus seems also to have applied to some of his successors, for example Nero, as we mentioned before. It is our aim now to show how Hadrian decided to add himself to the existing sacred space, and this will help us to situate the Pantheon in its context.

The main additions of Hadrian to the Campus Martius were the reconstruction of the Pantheon and the building of Hadrian's mausoleum (the present-day Castel S. Angelo). As far as the Pantheon is concerned, there is no doubt that its facade points to the Mausoleum of Augustus, establishing a direct visual link between the two (it may be that the original, northerly orientation of Agrippa's Pantheon was already exactly the same, and in this case Hadrian simply respected this orientation). Curiously, it is worth mentioning that there exists a further link – which seems, however, due to purely practical reasons – between the two monuments. On the limestone slabs which form the pavement in front of the Mausoleum, carefully etched lines may be recognised. These clearly refer to the design of the front of a huge temple which has dimensions very similar to those of the Pantheon. The standard explanation for this is that the (yet unworked) stones of the Pantheon's facade were disembarked from the river in the area in front of the Mausoleum and then prepared for assemblage before sending them to the building site, although, of course, one might at least suspect that Hadrian's architect reconstructed a pre-existing facade, planned at the same time as Augustus's Mausoleum (Inglese 2008).

Hadrian's Pantheon thus explicitly faced Augustus's funerary monument. As we have seen, at least according to the thesis defended in this paper, the Pantheon was the symbol of the emperor's power, identified with that of Rome itself. Clearly, then, the position of Hadrian's Mausoleum also had to be chosen to create a symbolic link to the monument. We propose that it was probably for this reason, despite there being at that time plenty of space in the Campus Martius proper, that Hadrian's Mausoleum was located on the opposite bank of the Tiber, to the north-west of the Pantheon (requiring, among other technical difficulties, the construction of a new bridge).

The centre of the monument is located at an azimuth of approximately 300° with respect to the facade of the Pantheon (Fig. 13). The azimuth of the setting sun at the summer solstice in AD 120 was about 302° in Rome. However, immediately behind the Castel S. Angelo there is Monte Mario, a hill some 100m high and 3km from the Pantheon. Taking the hill into account, we obtain a true azimuth for the setting sun at the summer solstice very close to 300°. Thus the sun at the summer solstice was seen to set over Hadrian's Mausoleum from the area in front of the Pantheon.[11] Interestingly, if we look at the opposite direction from the south-west, we see that the line crosses (some 7km distant) over the third largest mausoleum of Roman times (after Hadrian's and Augustus's), the so-called *Monte del Grano.* According to dated brickstamps, this tomb was constructed during the middle of the 2nd century AD. The original owner is unknown, but it was probably re-used as the tomb of the emperor Alexander Severus (AD 222-235).

All in all, it is likely that the Mausoleum of Hadrian was symbolically linked to the sacred space of the Campus Martius via the summer solstice sunset line. This line, however, of course imposed only one constraint on the position of the centre of the building. The second constraint may have come from the location of the Ara Pacis. Indeed the original position of this monument is reported to have been under a building at the east angle of the Piazza di S. Lorenzo in Lucina. It is then seen that this point and the centre of the Castle S. Angelo lie practically on the same parallel. Therefore, Hadrian's Mausoleum is located at the intersection between the summer solstice sunset line from the Pantheon and the east-west line from the Ara Pacis.

**Conclusion**

The relationship between the sacred space of the Campus Martius, the establishment of the emperor's cult and the connection of his power with the Sun God was initiated by Augustus. In doing so, he ordered the huge obelisk – together with its companion, erected in the Circus Maximus - to be transported there from the most ancient and important Egyptian temple of the sun, Heliopolis. The dedication of the two monuments (CIL 6:702) makes it clear that the process was already at an advanced state (Rehak and Younger 2006). Indeed, it reads "Imperator Caesar, son of the Divine [Julius], Augustus, Pontifex Maximus, Imperator for the 12th time, consul for the 11th time, with Tribunician power for the 14th time, with Egypt to the power of the Roman People subjected, gave this as a gift to the Sun". Further, and interestingly, from the imperial titles we obtain a date for the erection of the obelisk in Rome at 9 BC. This coincides with the start of the reformed calendar, established by Augustus by correcting an egregious error made in the original application of the Julian calendar after 45 BC, namely the addition of leap days not every fourth year, but every third (Hannah 2005: 98–122). Therefore, the symbolic function of the monument was also to credit the ruler as the "keeper" of the regular course of time. Clear traces of continuity in these ideas can be seen in Nero's time, and it has been our aim here to show that they can finally be seen "in action" in the Hadrian's Pantheon and in its visual and astronomical relationship with the other most representative monument of Hadrian in Campus Martius, his Mausoleum.[12]

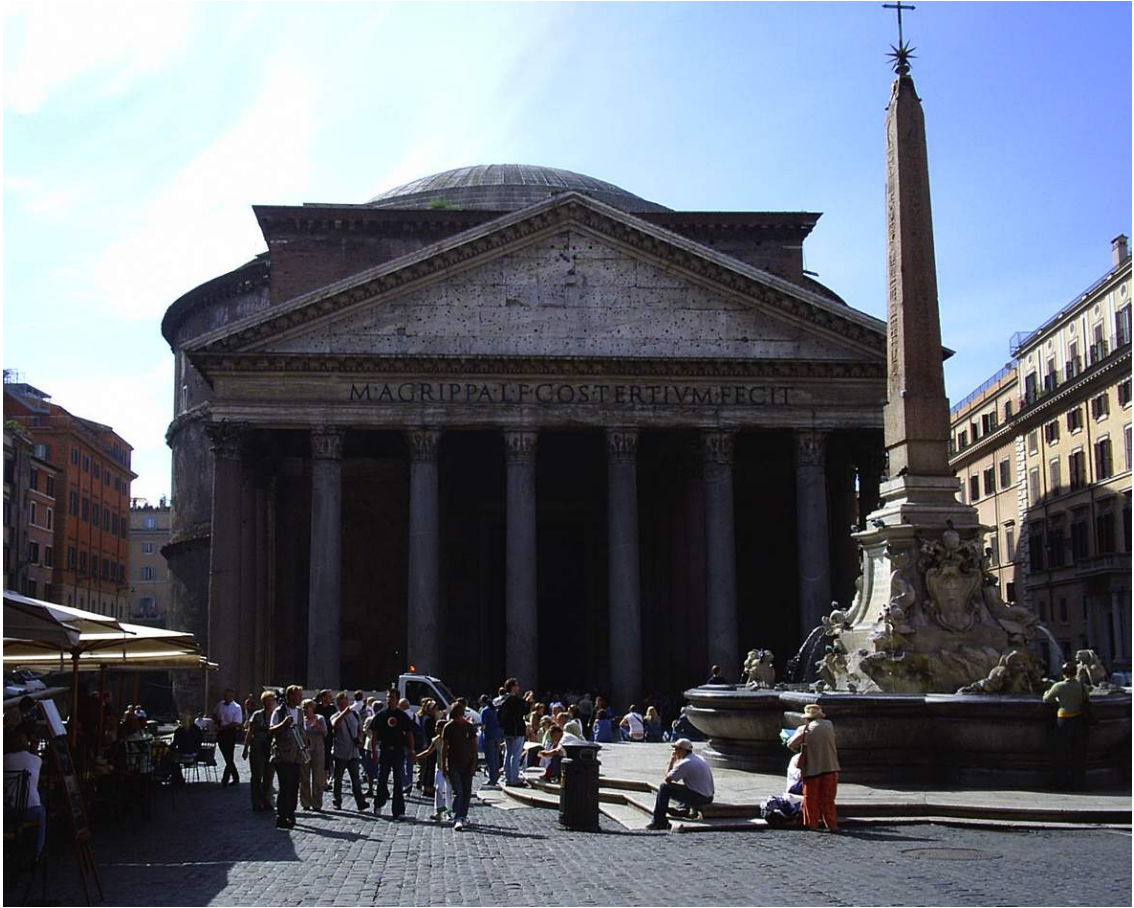

Fig. 1. Rome, the Pantheon: exterior.

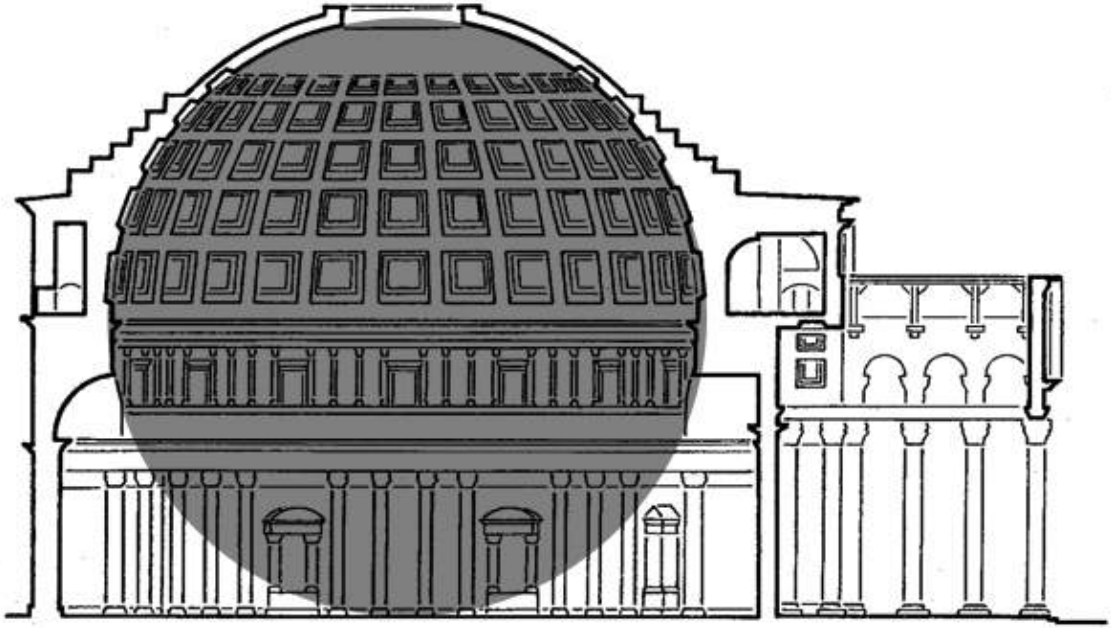

Fig. 2. North-south section through the Pantheon (north to the right), demonstrating hypothetical interior sphere.

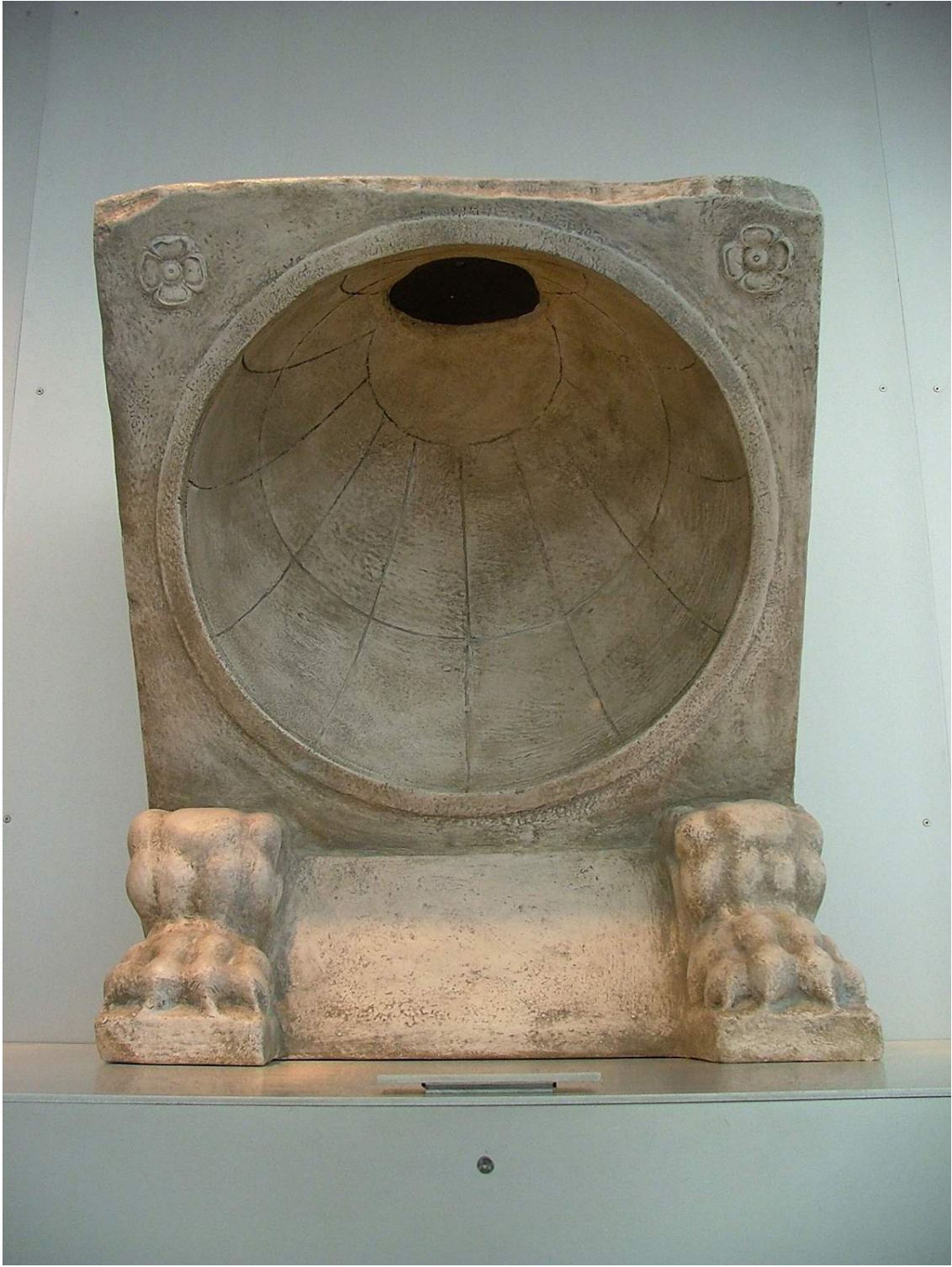

Fig. 3. Plaster cast of a roofed spherical sundial, Baelo Claudia, Spain.

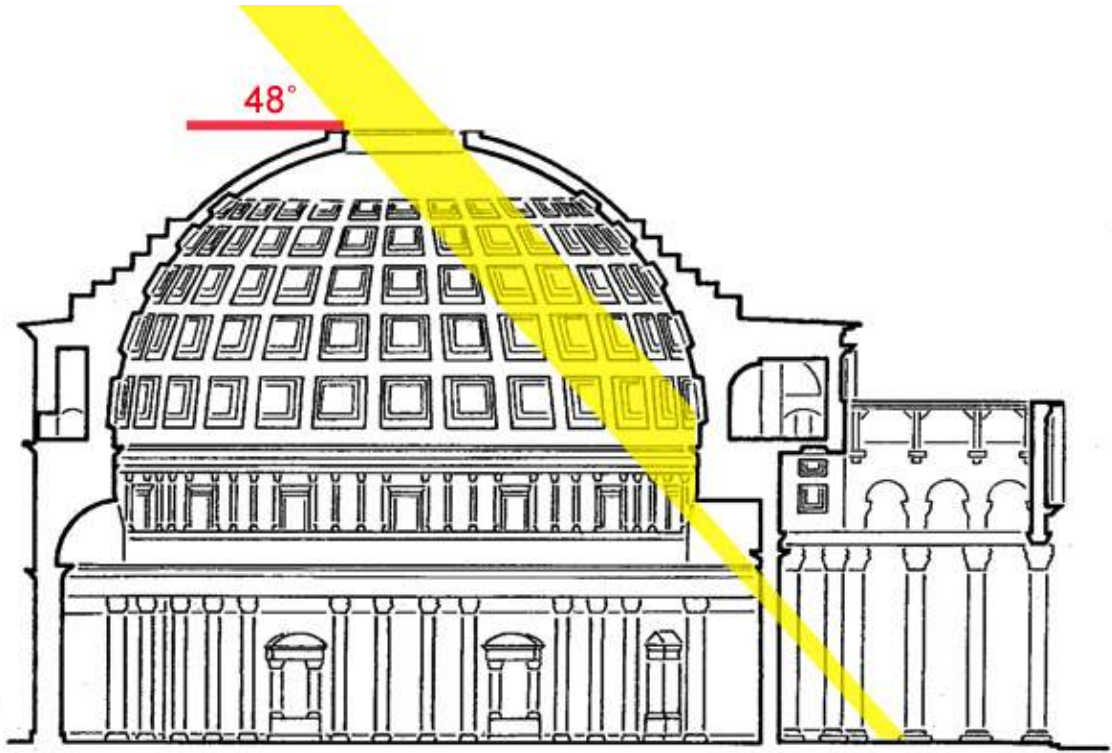

Fig. 4. Section through the Pantheon at the equinoxes, showing the fall of the noon sunlight, when the sun is at altitude 48˚.

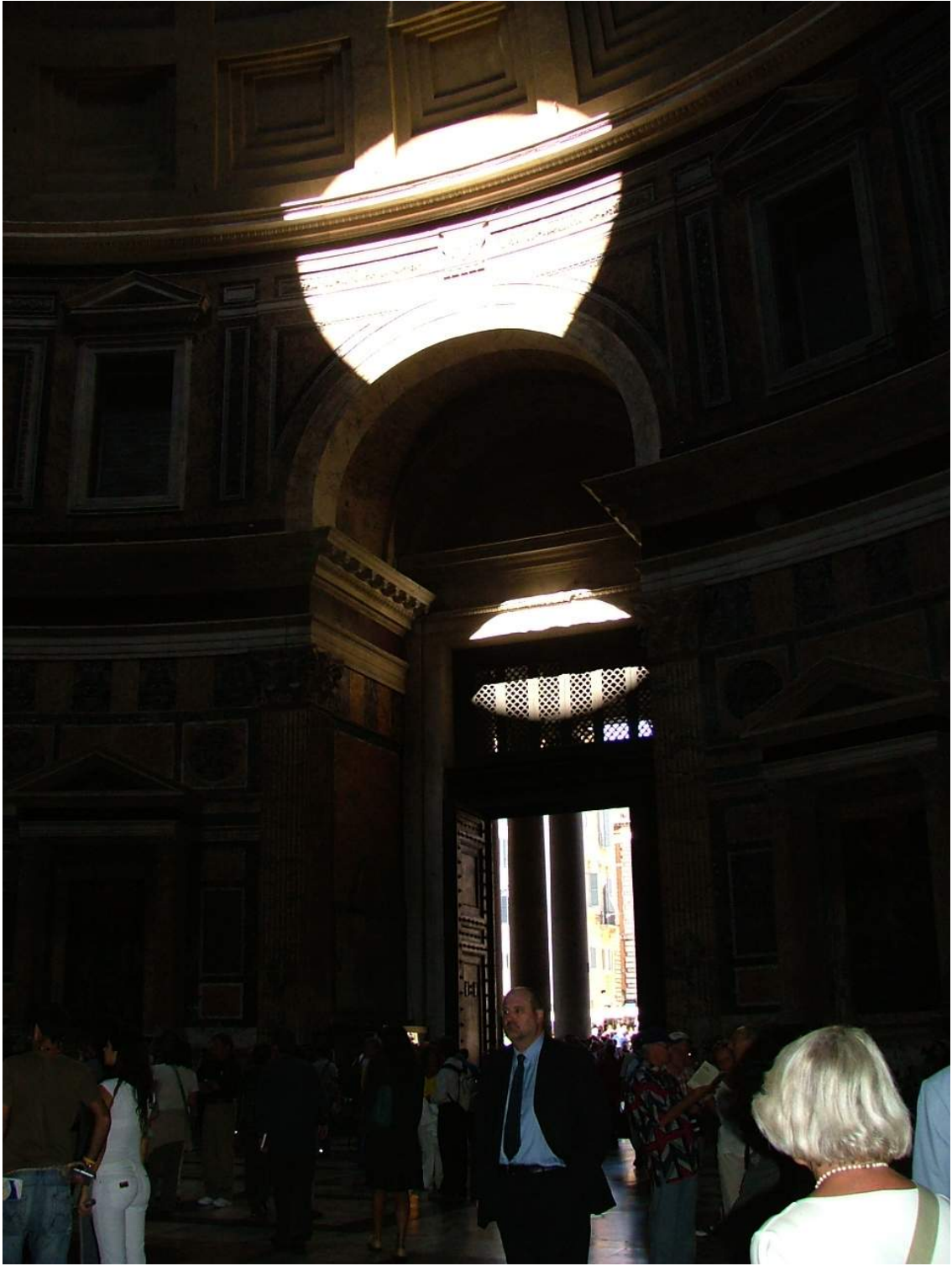

Fig. 5. Sunlight falling above the entrance of the Pantheon at local noon at the autumn equinox, 23 September 2005.

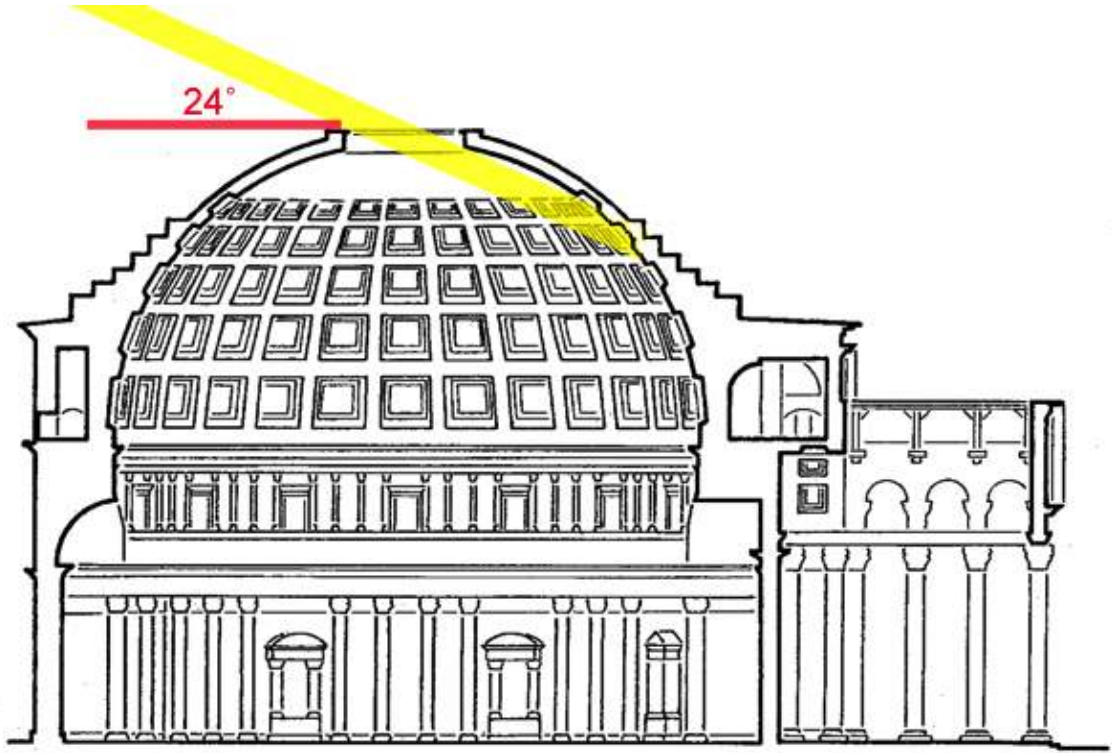

Fig. 6. Section through the Pantheon, showing the fall of the noon sunlight at the winter solstice, when the sun is at altitude 24˚.

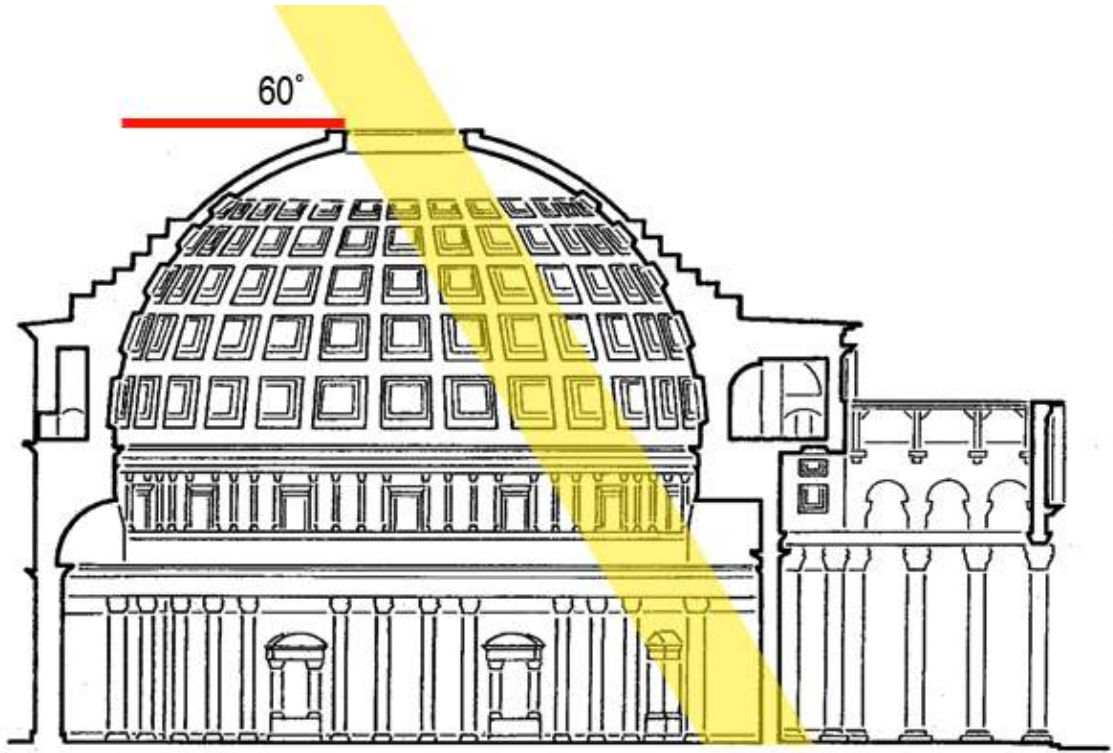

Fig. 7. Section through the Pantheon, showing the fall of the noon sunlight on 21 April.

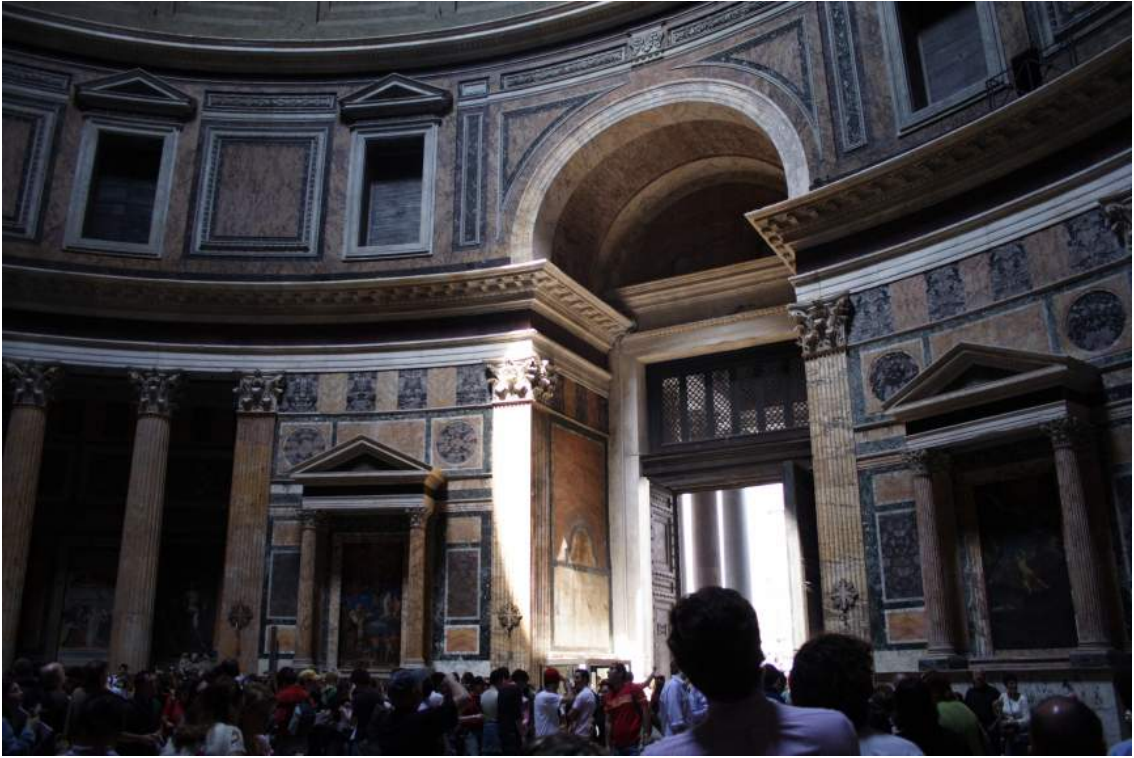

Fig. 8. Sunlight falling on the entrance of the Pantheon at local noon, 21 April 2007.

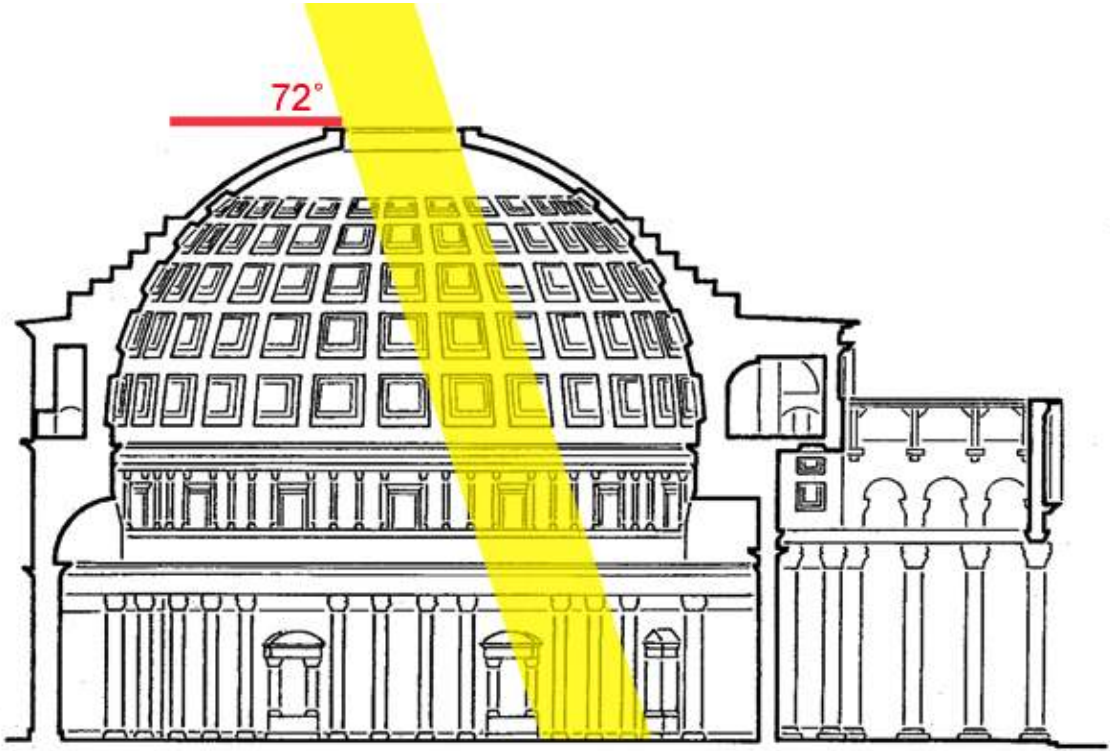

Fig. 9. Section through the Pantheon, showing the fall of the noon sunlight at the summer solstice, when the sun is at altitude 72˚.

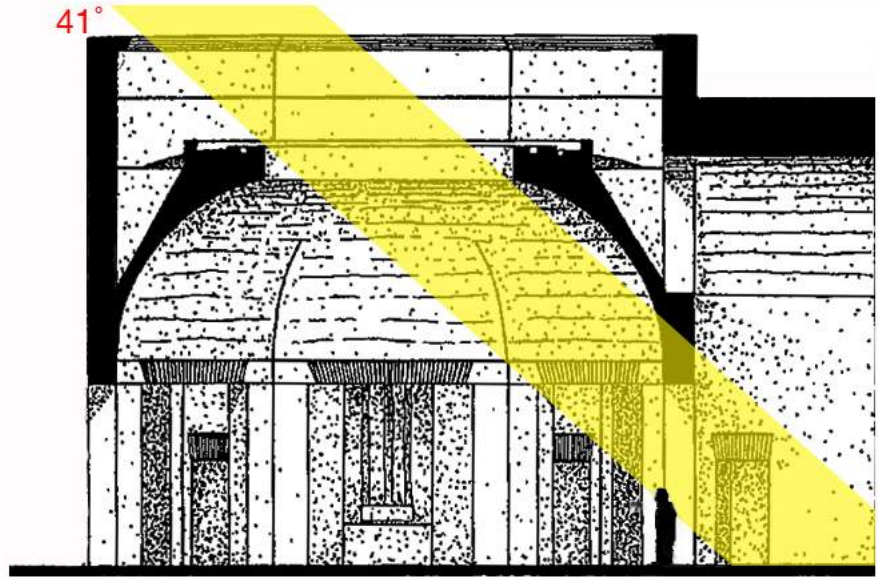

Fig. 10. North-south section through the Octagonal Room of the Golden House of Nero in Rome (north to the right), showing the fall of the noon sunlight on 13 October, when the sun is at altitude 41˚.

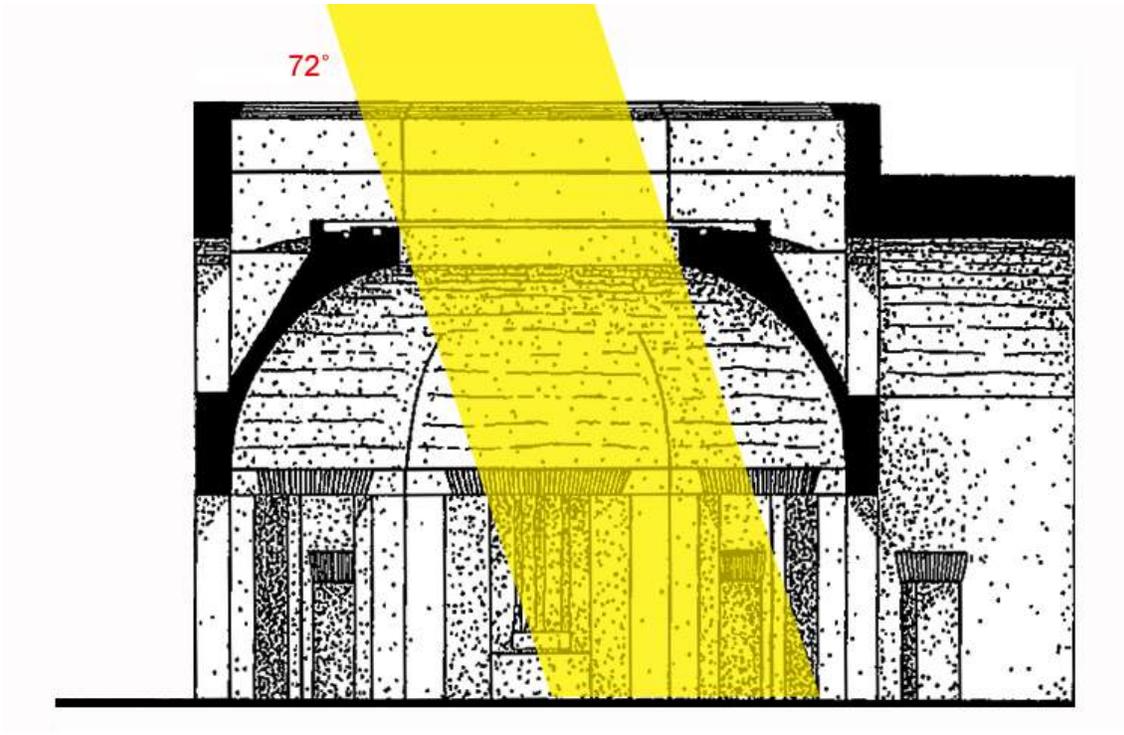

Fig. 11.  Section through the Octagonal Room of the Golden House of Nero in Rome, showing the fall of the noon sunlight at the summer solstice, when the sun is at altitude 72˚.

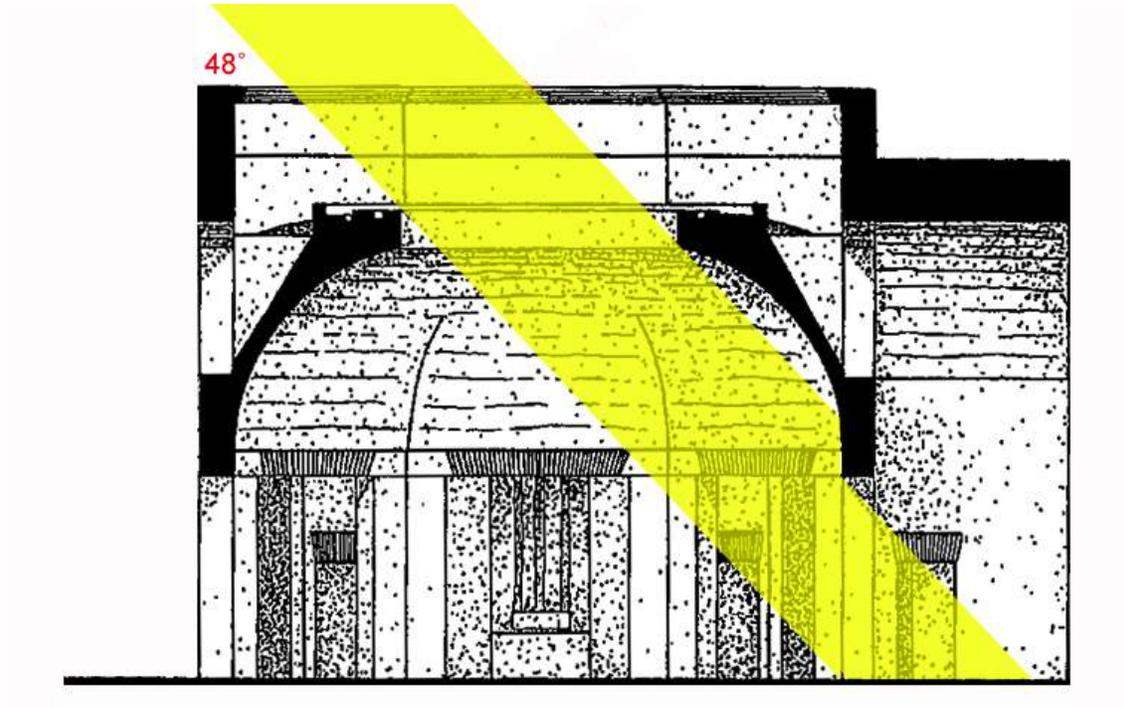

Fig. 12. Section through the Octagonal Room of the Golden House of Nero in Rome, showing the fall of the noon sunlight at the equinoxes, when the sun is at altitude 48˚.

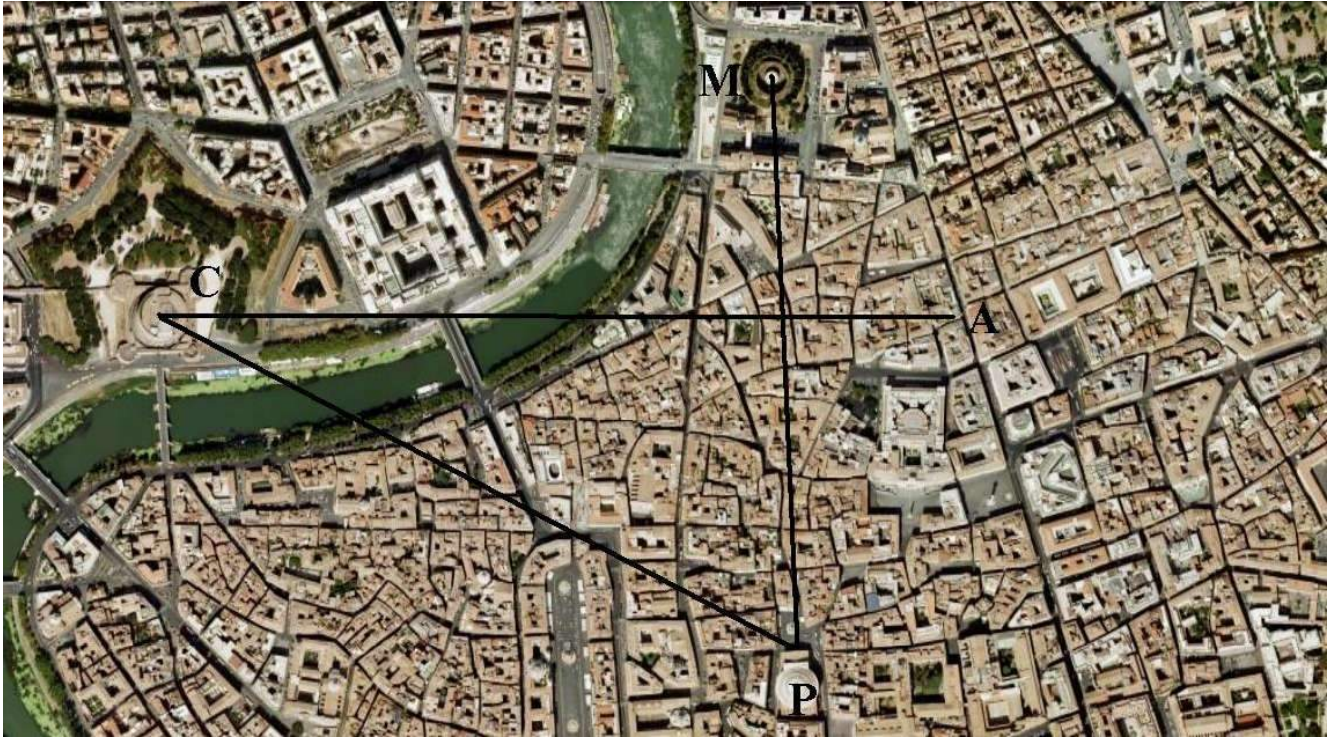

Fig. 13. A satellite image (north on top; adapted from Googlearth digital archive) showing the relationships between the Pantheon (P) the Mausoleum of Augustus (M) the Mausoleum of Hadrian (C) and the Ara Pacis (approximate original location A). The line P-M bears an azimuth of 354.5° and shows the axis of the Pantheon's entrance oriented towards Augustus' Mausoleum; the line P-C bears an azimuth of 300° and corresponds to the summer solstice sun setting behind S. Angelo Castle as viewed from the Pantheon; the line C-A should have had an azimuth very close to 270°.

**Notes**

1 Hetland argues controversially that the rebuilding was begun perhaps soon after AD 110 under Trajan, rather than entirely under Hadrian. De Fine Licht 1968 remains a fundamental study of the building; see also MacDonald 1976 and most recently Waddell 2008.

2 Nissen 1873 remains fundamental regarding the orientation of Roman temples. Wilson Jones 2003: 179 usefully rehearses the reasons why the Pantheon is not obviously a temple.

3 This datum is already present in the literature. Nevertheless, we have controlled it with a precision magnetic compass (Suunto Tandem, nominal precision ½ degree), taking into account magnetic deviation; the same holds for the data given in section 5.

4 Similarly, we did not find any explicit reference to the Moon cycle such as, for instance, a signpost for the northern maximal standstill which at that time corresponded to a moon altitude of around 77° at culmination. The coffering of the ceiling into twenty-eight parts has been thought to reflect the lunar cycle, but as Wilson Jones (2003: 241 n. 30) points out, twenty-nine would be a better approximation.

5 Contrast the above with de Fine Licht 1968: 295 n. 43, who saw no influence on the design of the Pantheon from the height of the sun at the solstices and the equinoxes.

6 The drawing of the Domus Aurea section in Figs. 10–12 is hypothetical, based on the corresponding east-west section commonly published; to our knowledge, no north-south section drawing exists, probably because of the steep and friable nature of the roofing (we are grateful to Professor Larry Bell for this information). Nonetheless, the symmetrical nature of the room implies that the two sections should be more or less the same.

7. *. . . praecipua cenationum rotunda, quae perpetuo diebus ac noctibus vice mundi circumageretur . . .*

8 Note, however, the proposed identification of the *coenatio rotunda* with an extraordinary structure recently excavated on the *Caelian* hill: http://www.beniculturali.it/mibac/export/MiBAC/sito-MiBAC/Contenuti/Ministero/UfficioStampa/ComunicatiStampa/visualizza_asset.html_1690133015.html (accessed 29 September 2009).

9 It is assumed here, along with Ramsey and Licht (1997: 8-12, 19-57) that the *ludi Veneris Genetricis* (Games for Venus Genetrix) were moved from their original date in September to 20-30 July by 44 BC, and at some stage were renamed the *ludi Victoriae Caesaris* (Games for the Victory of Caesar). See also Hannah 1998.

10 *Ei, qui est in campo, divus Augustus addidit mirabilem usum ad deprendendas solis umbras dierumque ac noctium ita magnitudines, strato lapide ad longitudinem obelisci, cui par fieret umbra brumae confectae die sexta hora paulatimque per regulas, quae sunt ex aere inclusae, singulis diebus decresceret ac rursus augeresceret, digna cognitu res, ingenio Facundi Novi mathematici. is apici auratam pilam addidit, cuius vertice umbra colligeretur in se ipsam, alias enormiter iaculante apice, ratione, ut ferunt, a capite hominis intellecta.*

11 At that time to the north-west of the Pantheon there was a building, Nero's Baths, whose precise perimeter and elevation are unknown. Although it is unlikely, we cannot exclude the possibility that it might have partially obstructed the direct view to the Mausoleum from the Pronaos.

12 Clearly, this is not the place for a thorough-going attempt to gain more insight into the implications that these facts may have in the problem of interpreting Hadrian's thought and "philosophy of power". It seems likely, however, that at least part of the project originated from the emperor's own interest and nostalgia for Egypt. Egyptian resemblances are to be seen explicitly in the *Villa Adriana*, the emperor's retreat in Tivoli, where several architectural elements clearly recall Egypt (Roullet 1972) and where one of the most enigmatic buildings, the so-called "tower" of Roccabruna, a squared-plan edifice incorporating an octagonal room, is oriented to the summer solstice sunset (Castellani 2005). Finally, the recently discovered cenotaph of Antinous, which contains explicit references to the Egyptian world (Mari and Sgalambro 2007), opens up even further possibilities for research in this direction.